\documentclass[letter]{aa} 

\usepackage{graphicx}
\usepackage{soul}
\usepackage{ulem}
\usepackage{txfonts}
\begin{document}

   \title{Soft Gamma-Ray Spectral and Time evolution of the GRB 221009A: prompt and afterglow emission with \textbf{\textit{INTEGRAL}/\textrm{IBIS-PICsIT}}}

   \author{James Rodi
          \inst{1}
          \and
          Pietro Ubertini
          \inst{1}
          }

   \institute{INAF - Istituto di Astrofisica e Planetologia Spaziali,
              via Fosso del Cavaliere 100; 00133 Roma, Italy\\
              \email{james.rodi@inaf.it}
             \\
             \email{pietro.ubertini@inaf.it}
             }

   \date{Received -; accepted -}

   \titlerunning{Spectral and Time Evolution of GRB 221009A with IBIS-PICsIT}
    \authorrunning{Rodi and Ubertini}

 
  \abstract
   {}
   {The gamma-ray burst (GRB) 221009A, with its extreme brightness, has provided the opportunity to explore GRB prompt and afterglow emission behavior on short time scales with high statistics.  In conjunction with detection up to very high-energy gamma-rays, studies of this event shed light on the emission processes at work in the initial phases of GRBs emission.}
   {Using \textit{INTEGRAL}/\textrm{IBIS}'s soft gamma-ray detector, \textrm{PICsIT} (\(200-2600\) keV), we studied the temporal and spectral evolution during the prompt phase and the early afterglow period.}
   {We found a "flux-tracking" behavior with the source spectrum "softer" when brighter.  However the relationship between the spectral index and the flux changes during the burst.  The \textrm{PICsIT} light curve shows afterglow emission begins to dominate at \( \sim T_0 +630\) s and decays with a slope of \(1.6 \pm 0.2\), consistent with the slopes reported at soft X-rays.}
   {}

   \keywords{Gamma rays: general --
                (stars:) gamma-ray burst: general --
                (stars:) gamma-ray burst: individual : (GRB 221009A)
               }
   \maketitle

%

\section{Introduction}

The long gamma-ray burst (GRB) GRB 221009A was likely the brightest GRB ever detected \citep{burns2023}.  The observation of the prompt emission was initially reported by \textit{Fermi}/\textrm{GBM} (\(T_0 = \) 13:17:00 UTC) \citep{veres2022}.  In view of the extreme GRB flux, spanning from low energy X-rays to very high gamma-rays, detections were reported by numerous instruments, including detectors not built to detect GRBs like \textrm{GAIA}, \textrm{SOHO} \citep{esa2022}, \textrm{Solar Orbiter} \citep{xiao2022}, \textit{CSES} \textrm{HPPL} \citep{battiston2023} and others.  Most of the GRB devoted telescopes and observatories were obviously triggered, many of them with severe problems from detector dead-time, pile-up and telemetry saturation: \textit{Swift}/\textrm{BAT} \citep{dichiara2022}; \textrm{MAXI} \citep{negaro2022}; \textit{Fermi}/\textrm{LAT} \citep{bissaldi2022}; \textit{AGILE} \citep{piano2022,ursi2022}; \textit{INTEGRAL} \citep{gotz2022}; \textit{Konus-Wind} \citep{frederiks2022}; \textit{Insight-HMXT} \citep{tan2022}; \textit{STPSat-6}/\textrm{SIRI} \citep{mitchell2022}; \textit{GECAM} \citep{liu2022}; \textit{SRG}/\textrm{ART-XC} \citep{lapshov2022}; \textit{GRBAlpha} \citep{ripa2022}).  None the less, the usable data from the instruments will enable studies of GRB prompt emission evolution at high time resolution and high statistics.  Combined with multi-wavelength afterglow detections up to TeV energies, GRB 221009A is a unique opportunity to explore numerous aspects of GRB behavior with particular regard to the initial phase of the transition from prompt to afterglow gamma-ray emission. 

In this work, we analyse the soft gamma-ray evolution of GRB 221009A in the \(200-2600\) keV energy range using the spectral-timing data provided by the \textrm{IBIS/PICsIT} gamma-ray telescope aboard \textit{INTEGRAL} to study how the prompt emission varies throughout the burst.  Additionally, we explored the characteristics of the afterglow emission during the prompt phase and shortly after it.

\section{Observations and analysis}

The \textit{INTErnational Gamma-Ray Astrophysics Laboratory} (\textit{INTEGRAL}) was launched in October 2002 from Baikonur, Kazachstan \citep{jensen2003} in a high Earth orbit providing long un-interrrupted observations due to the \( \sim 2.5\)-day elliptical orbit, resulting in an all-sky coverage for more than of 85\% of the operation time.  With its on-board suite of instruments, \textit{INTEGRAL} \citep{winkler2003} spans 3 keV \(-\) 10 MeV with fields-of-view (FoV) ranging from \( \sim 100 - 1000\) deg\(^{2}\).  Also, the active veto shields of \textrm{SPI} \citep{vedrenne2003} and \textrm{IBIS} \citep{ubertini2003} feature large sensitive exposed area to high energy photons and can detect impulsive events outside the FoVs of the imaging instruments with high sensitivity \citep{vonkienlin2003,savchenko2017}.  

\textit{INTEGRAL} was observing XTE J\(1701-462\) at the trigger time of GRB 221009A (2022-10-09 13:17:00.0 UTC) with an angle of \( \sim 65.8^{\circ}\) off-axis from the pointing direction.  The prompt emission lasted more than 600s, including the small precursor peak and spanned \textit{INTEGRAL} observations \(255800290010-255800300010\).  Beginning at 2022-10-10 13:27:56 UTC, \textit{INTEGRAL} began pointed observations of the afterglow emission.  Analysis and interpretation of these afterglow observations are covered in \cite{savchenko2023}. 
While the \textrm{SPI-ACS} has a higher sensitivity as a GRB monitor due to its large effective area (0.7 m\(^2\)), it has only a single energy channel (\(> 75 \) keV \citep{vonkienlin2003}).

\textrm{PICsIT}, \textrm{IBIS}'s soft gamma-ray detector, covers energies from 170 keV to 10 MeV with two commonly used data types: spectral-timing and spectral-imaging \citep{labanti2003}.  The spectral-timing data has 7.8-ms time resolution in 8 broad energy bands (\(200-2600\) keV).  This data type sums all the counts in all the detector pixels and thus does not have any position resolution.  In contrast, the spectral-imaging data type sums the counts in each pixel over the span of an \textit{INTEGRAL} pointing (\( \sim 1800\) s).  Using the spectral-timing data, \textrm{PICsIT} is able to detect impulsive events both inside and outside the FoV \citep{bianchin2009,bianchin2011,savchenko2017}, though with reduced sensitivity for those outside to shielding of the \textrm{IBIS} walls.  It is possible to account for the absorbing materials to also produce spectra \citep{bianchin2009,rodi2021}.  For the above mentioned reasons, in this paper we focus on \textrm{IBIS-PICsIT} data which provides 8 broad energy channels and a higher time resolution (7.8-ms vs 50-ms of \textrm{SPI-ACS}).  We analyzed the observations \(255800290010-255800300010\) as part of an accepted \textit{INTEGRAL} proposal to analyze the \textrm{PICsIT} spectral-timing data for reported GRBs.

\textit{INTEGRAL} telemeters its data to the ground in real-time \citep{winkler2003}.  However, the size of the on-board buffer was optimized for 'standard' observations of fields like the Galaxy Centre, Crab etc, and thus is not ideal to cope with impulsive, high-flux events, due to the limited bandwidth available at the time of the satellite design and construction.  Thus periods of very high count-rate (e.g. GRBs in the FoV) can result in buffer overflows and data gaps for on-board instruments.  To remove such bad time intervals (BTIs), we removed periods with 3 or more time bins missing.  Additionally we searched for detector saturation.  No time bins were found at or near the maximum possible value allowed by the dedicated telemetry space, indicating that \textrm{PICsIT} did not suffer from pile-up effects during the burst.

\section{Results} \label{sec:res}

\subsection{Temporal evolution}

\begin{figure*}[h!]
  \centering
  \includegraphics[scale=0.7, angle=180,trim = 15mm 40mm 60mm 70mm, clip]{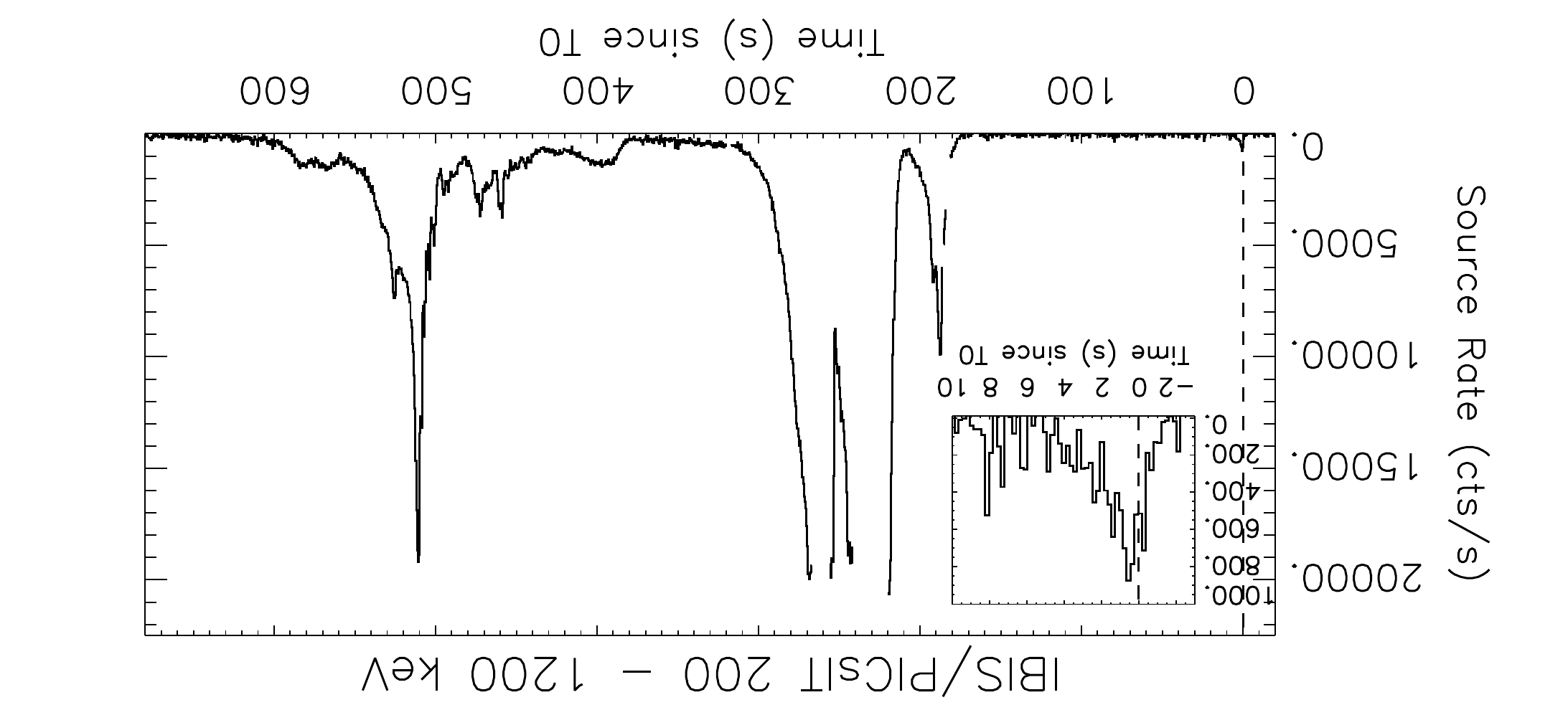} 
\caption{Time evolution of the background subtracted \textrm{PICsIT} light curve of GRB 202909A in the energy range \(200-1200\) keV. Each time bin is integrated over 500-ms to increase the statistics and avoid empty time bins. The dashed line corresponds \(T_0=\) 13:17:00 UTC, when the precursor has been clearly detected by \textrm{IBIS-PICsIT}. The inset shows the precursor with 200-ms time resolution.}
    \label{fig:lc_long}
\end{figure*}

GRB 221009A began with a precursor at 13:17:00 UTC.  Figure~\ref{fig:lc_long} shows the \(200-1200\) keV \textrm{PICsIT} light curve on a 500-ms time resolution.  A dashed line denotes \(T_0\) (\(T_0 =\) 13:17:00 UTC).  The precursor is shown in an inset to better show the behavior.  The feature shows a fast rise with an exponential decay lasting for approximately 7 s.  No other significant emission was detected in \textrm{PICsIT} until \( \sim T_0 + 177\) s.  Subsequently, Pulse 1 (\( \sim T_0 + 177 - 205\) s) starts, peaks around 11,000 cts/s, decays to \( \sim 5,000 \) cts/s after which a sub-flare begins (peak \( \sim 7,000\) cts/s) then further decays to a low inter-pulse flux at \( \sim T_0 + 210 \) s. 

Then Pulse 2 (\(\sim T_0 + 210 - 252\) s) begins, increasing from an inter-pulse level of \( \sim 1000\) cts/s to \( \sim 21,000\) cts/s in the span of \( \sim 8\) s, after which the \textrm{PICsIT} data suffers from telemetry issues from \( \sim T_0 + 220\) s.  Thus the peak of the flare is not detected with \textrm{PICsIT}.  The gap ends at approximately \( T_0 + 243\) s with a \textrm{PICsIT} count-rate of \( \sim 19,000\) cts/s.  After which the GRB flux rapidly decreases to \( \sim 15,000\) cts/s before a more gradual decrease until roughly \(T_0 + 252\) s. 

\begin{figure*}[h!]
  \centering
  \includegraphics[scale=0.77, angle=180,trim = 5mm 10mm 45mm 0mm, clip]{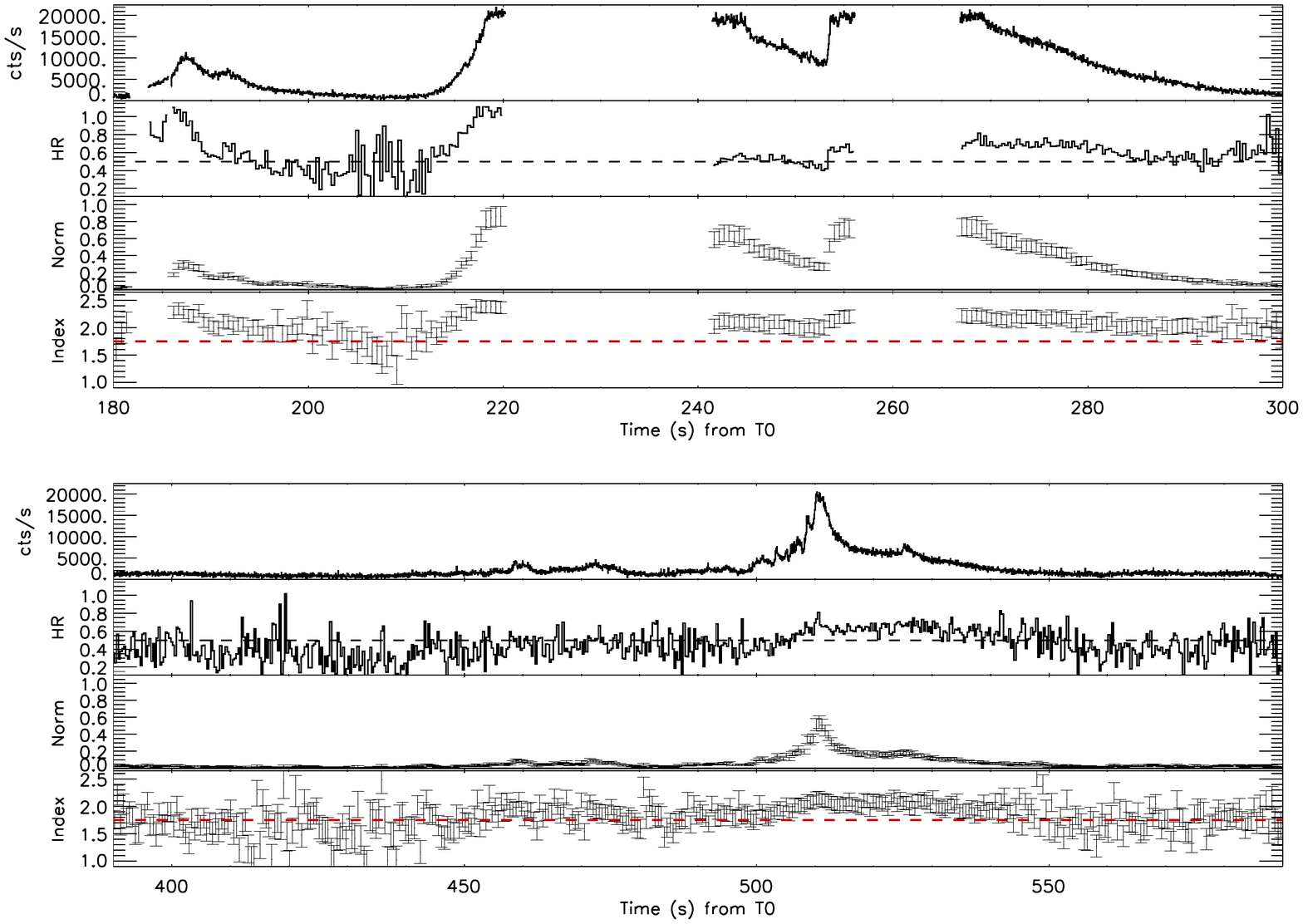} 
\caption{GRB 221009A \(200-1200\) keV \textrm{PICsIT} light curve with the corresponding hardness ratio HR=(\(200-312 \) keV)/(\(570-1200\) keV), normalization at 300 keV (ph/cm\(^2\)/s/keV), and photon index to power-law fits.  The black dashed line denotes HR\(=0.5\) for reference, and the red dashed line at an index of 1.75 for reference.}
    \label{fig:lc_med}
\end{figure*}

Pulse 3 (\(\sim T_0 + 252 - 320 \) s) begins with a fast increase (\( \sim 8,500\) to \( \sim 20,00\) cts/s).  Much of the Pulse 3 rise overlaps with the decay of Pulse 2 and thus is not observed. 
 \textrm{PICsIT} again has telemetry issues at \( \sim T_0 + 255\) s and is unable to monitor the peak behavior until \( T_0 + 268 \)s when the flux begins to decrease from a level of \( \sim 19,500\) cts/s to a few 100 cts/s.  Followed by an inter-pulse period until approximately \( T_0 + 380\) s.  

Pulse 4 (\(\sim T_0 + 380 - 600\) s) commences with a comparatively low flux shoulder-like feature that slowly varies between \( \sim 500 - 2,000\) cts/s until roughly \( T_0 + 495 \) s.  On top of this feature several sub-flares occur, lasting \( \sim 5-10 \) s with count-rates near 4,000 cts/s at their peaks.  After \( \sim T_0 + 495 \) s, the flux dramatically increases to a peak of approximately 20,000 cts/s in \( \sim 15 \) s with multiple sub-flares present during the rise.  The Pulse peaks before the flux decreases to roughly 10,000 cts/s in approximately 5 s and then decaying gradually until \( \sim T_0 + 550\) s after which the shoulder-like feature continues for roughly 50 s more.  The results after \( T_0 + 1000 \) s are discussed in \cite{savchenko2023} as part of the INTEGRAL afterglow follow-up observations.  
 
We investigated the hardness ratio (HR) evolution, defined as count rate (\(200- 312\) keV/\(570-1200\) keV), during the four pulses.  Their evolution and corresponding HR are shown in Figure~\ref{fig:lc_med}, where a dashed line is drawn at HR \( =0.5\) for reference.  The evolution of the normalization at 300 keV and the photon index (\( \Gamma\)) are also shown and will be discussed in detail in Section~\ref{sec:spec_ev}. GRB 221009A shows an evolution of hardness throughout the pulses with generally higher HR values ("softer") at higher fluxes and lower HR values ("harder") at lower fluxes. However some variability is present during the inter-pulse periods due to the relatively low count-rate possibly generating statistical fluctuations.

For Pulse 1, much of the data during the rise phase are missing due to a telemetry gap.  Thus the existing data show predominately the peak and decay behavior with the hardness evolving from \( \sim 1.1 \) at the peak to \( \sim 0.3 \) just before the inter-pulse phase.  

Pulse 2 shows a more complicated behavior.  During the rise, the hardness begins at roughly 0.5 and increases to nearly 1.2 as the flux increases before the data gap.  In contrast, the dip shows a near constant hardness at \( \sim 0.5\) as the flux decreases followed by a sharp increase in both hardness and flux (the beginning of Pulse 3) before flattening at HR \( \sim 0.6\) prior to the next data gap.  When the decay phase of Pulse 3 starts, the hardness is approximately 0.7 and decreases to a value of roughly 0.5 before beginning the inter-pulse period between Pulses 3 and 4.

Pulse 4 behaves differently from the prior three.  Starting from \(T_0 +400\) s, the HR varies from \( \sim 0.2 - 0.4\).  After the HR increases from \( \sim 0.4\) to \( \sim 0.6\) and peaks at nearly 0.75 at the peak of the pulse.  During the decay phase, the hardness decreases to roughly 0.6 and remains nearly constant until \( T_0 +540 \)s when the HR continues to decrease.  However, there are large variations in the behavior as the flux decreases until the afterglow begins at roughly \( T_0 + 600\) s.


\subsection{Spectral evolution}
\label{sec:spec_ev}

Expanding on the HR evolution analysis, we fitted the \textrm{PICsIT} data in 7 channels spanning \(250 -2600 \) keV using \(0.5-\)s integration throughout to study the changes in the spectral parameters.  We found that the spectra are adequately fit with a power-law model and did not require a high-energy cutoff.  The evolutions of the normalization at 300 keV in ph/cm\(^2\)/s/keV and the photon index (\( \Gamma\)) are shown in Figure~\ref{fig:lc_med}, as mentioned above.

In agreement with the results from HR evolution, there is a general trend of increasing photon index with increasing flux.  Pulse 1 shows an increase from \( \Gamma \sim 1.7\) at \(\sim T_0 + 180\) s prior to the gap and indexes of approximately 2.3 at the peak of the pulse.  After which, \( \Gamma \) decreases to roughly 1.4 during the inter-pulse period at \( \sim T_0 + 207\) s.

Subsequently, the photon index during the Pulse 2 rise rapidly increases to \( \sim 2.4\) prior to the gap at \( \sim T_0 + 220\) s where the values flatten while the normalization shows an increase, though they are consistent with a constant value.  Post-gap \( \Gamma \) gradually decreases before increasing to \( \Gamma \sim 2.2\) during the rise of Pulse 3 when the next gap begins.  When the data restarts, the photon index is at a similar value and gradually decreases to roughly 2.  

Following the next inter-pulse period at \( \sim T_0 + 400\) s, \( \Gamma \sim 1.6\) until approximately \(T_0 + 450\) s, though a large amount of scatter due to the low flux.  Next, \( \Gamma\) slowly varies between \( \sim 1.6\) and \( \sim 1.9\) until roughly \(T_0 + 480\) s after which the photon index is constant at approximately 1.75 for nearly 20s.  Then as the flux increases rapidly, \( \Gamma \) increases to \( \sim 2.1\) at approximately \(T_0 + 511\) s when the pulse peaks.  During the decay phase, the normalization drops by a factor of roughly 3 with only a small decrease in photon index (\( \Gamma \sim 2\)).  After \( \sim T_0 + 515\) s the flux plateaus for roughly 10 s with photon index constant before decreasing exponentially until \(\sim T_0 + 560\) s while \( \Gamma \) decreases from roughly 1.9 to 1.5.  After this the normalization and photon index behave similarly to the period during \( \sim T_0 +400 - 450\) s.

\section{Discussion} \label{sec:dis}

\subsection{Prompt emission}

\begin{figure*}[h!]
  \centering
  \includegraphics[scale=0.6, angle=180,trim = 5mm 10mm 40mm 0mm, clip]{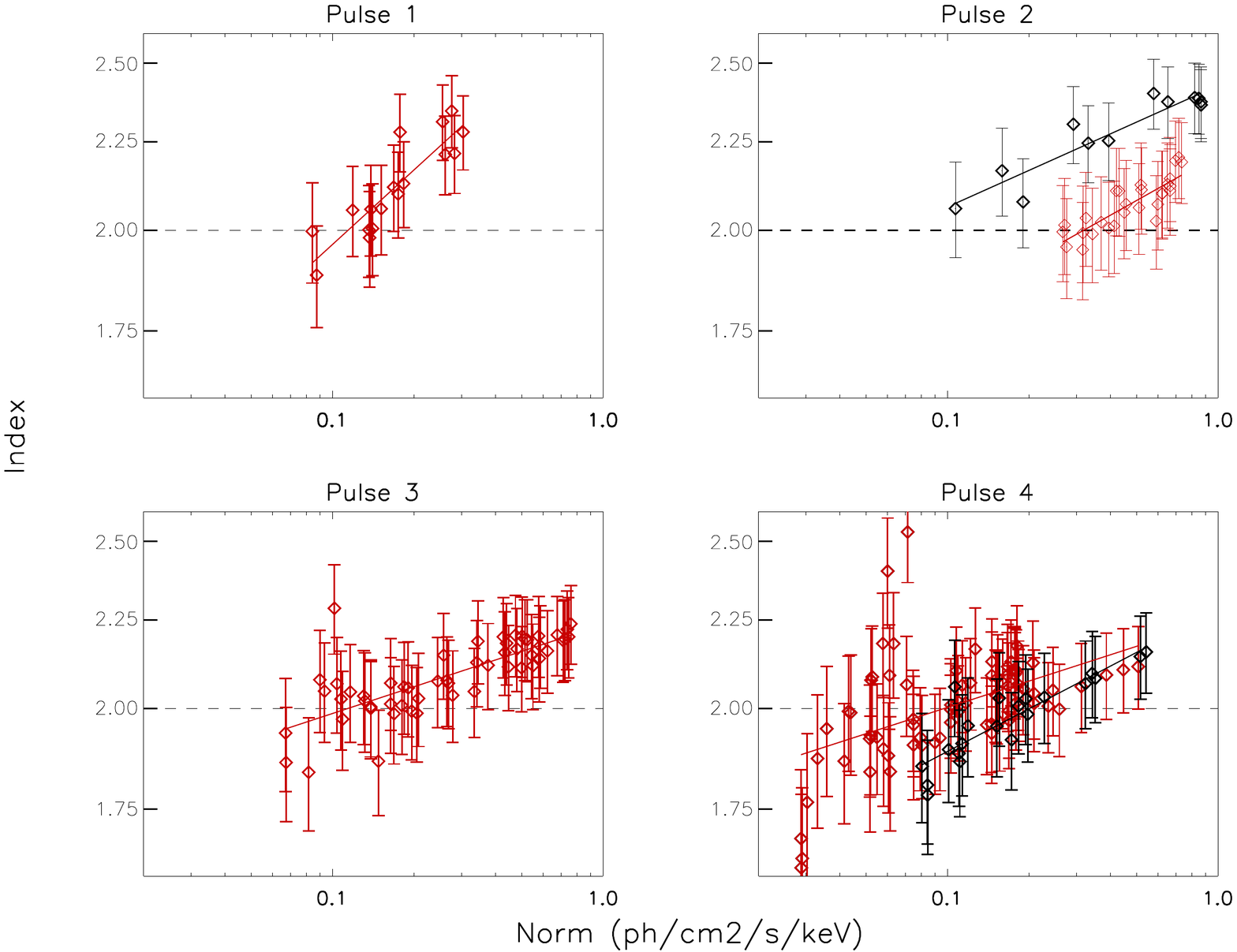} 
\caption{Evolution of the photon index with flux during the rise (black) and decay (red) for the four pulses. A dashed line is plotted at a photon index of 2 for reference.} 
    \label{fig:norm_gamma}
\end{figure*}

As seen in Figure~\ref{fig:lc_med}, the spectral behavior and flux show correlated behavior.  Similar behavior was reported by \cite{an2023} during from \(T_0 + 180 - 300\) s (though with a Band model) and has been found in \( \sim 2/3\) of multi-pulse GRBs studied by \cite{li2021}.  In the case of GRB 221009A, we found that the relationship between the flux and photon index changes throughout and between the pulses.  Figure~\ref{fig:norm_gamma} shows the different behavior in the rise and decay of the pulses with power-law fits to each phase.  

The fit to the Pulse 1 decay found a slope of \( 0.14 \pm 0.02\) and a normalization of \( 1.00 \pm 0.03\).  However, the lack of rise data means it is not possible to compare the two phases.  Pulse 2 shows a similar but significantly different slope between the rise and and decay with values of \( 0.07 \pm 0.01 \) and \( 0.09 \pm 0.01\), respectively, but the two have noticeably different normalizations (\( 0.89 \pm 0.01 \) vs \(0.79 \pm 0.01\)).  Both slopes are significantly lower than the slope and normalization of the Pulse 1 decay.

Due to the brevity of the Pulse 3 rise, we studied only the decay phase.  Its behavior is more complicated than the previous two pulses.  The fit has a slope of \( 0.051 \pm 0.006\) and a normalization of \( 0.08 \pm 0.02\), which is flatter than both the Pulses 1 and 2, but with a normalization comparable to the Phase 2 decay.  However, the  data show a possible flattening behavior below \( \sim 0.25 \) ph/cm\(^2\)/s/keV.  

The Pulse 4 behavior rise has a slope of \( 0.08 \pm 0.01\) and normalization of \(0.83 \pm 0.02\) comparable to Pulse 2.  In contrast, the Pulse 4 decay behavior is similar to the Pulse 3 decay (slope of \( 0.051 \pm 0.01\) and normalization of \(0.81 \pm 0.01\)) with the data below roughly 0.25 ph/cm\(^2\)/s/keV again apparently flattening to nearly constant.  Thus the spectral evolution of the pulses changes throughout the prompt emission, though some phases of the later pulses show similar behavior.  


\textrm{PICsIT}'s spectral-timing data's limited energy range makes direct comparisons with instruments that extend to lower energies difficult to compare if there is spectral curvature, as is reported by \textit{GRBAlpha} \citep{ripa2023}, \textit{Konus-Wind} \citep{frederiks2023}, \textit{GECAM} \citep{an2023}.  However, \textit{AGILE}/\textrm{MCAL} found the data from \(T_0 +181.00 - 194.03 \) s well fit by a power-law of \( \Gamma = 2.07 \pm 0.04\) \citep{ursi2022}.  While \textrm{PICsIT} does not have data for during the early portion of that time period, Figure~\ref{fig:lc_med} shows similar values.   

\cite{yang2023} performed a time-resolved spectral analysis of the prompt emission using \textit{GECAM-C} in the 6 keV \(-\) 6 MeV energy range.  They fit the spectra to a physical model assuming a synchrotron emission origin.  The \textrm{PICsIT} data have a  limited energy range that prevent us to perform the same data analysis, but we can compare the time-evolution trends with our results.


Anyway, the \cite{yang2023} results show a strong "flux-tracking" behavior for the power-law index of the injection rate (\(q\)) during Pulse 2.  (They do not present any results during Pulse 1.)  The behavior during Pulse 3 has an increase in \(q\) with flux, though the correlation is less clear than for Pulse 2.  Unfortunately, much of the lack of correlation occurs during the gap in the \textrm{PICsIT} data so a direct comparison is not possible.  Also, Pulse 3 analysis in \cite{yang2023} ends at \( \sim T_0 + 272\), and thus we are not able to compare the later Pulse 3 decay behavior that suggests little to no flux correlation.  During Pulse 4, no significant "flux-tracking" is present in \cite{yang2023}.  While \(q\) is highest during the peak, the errors are large during the results shown (\( \sim T_0 + 500 - 520\) s).  Overall, it is worth mentioning that the \textrm{PICsIT} results in Figure~\ref{fig:norm_gamma} are consistent with those from \cite{yang2023}.  
  Thus are data are consistent with their interpretation of a synchroton-emission origin from an expanding emission region and a decaying global magnetic field and therefor a jet which is Poynting-flux dominated.

\subsection{Afterglow emission}

\begin{figure*}[h!]
  \centering
  \includegraphics[scale=0.6, angle=180,trim = 5mm 10mm 30mm 60mm, clip]{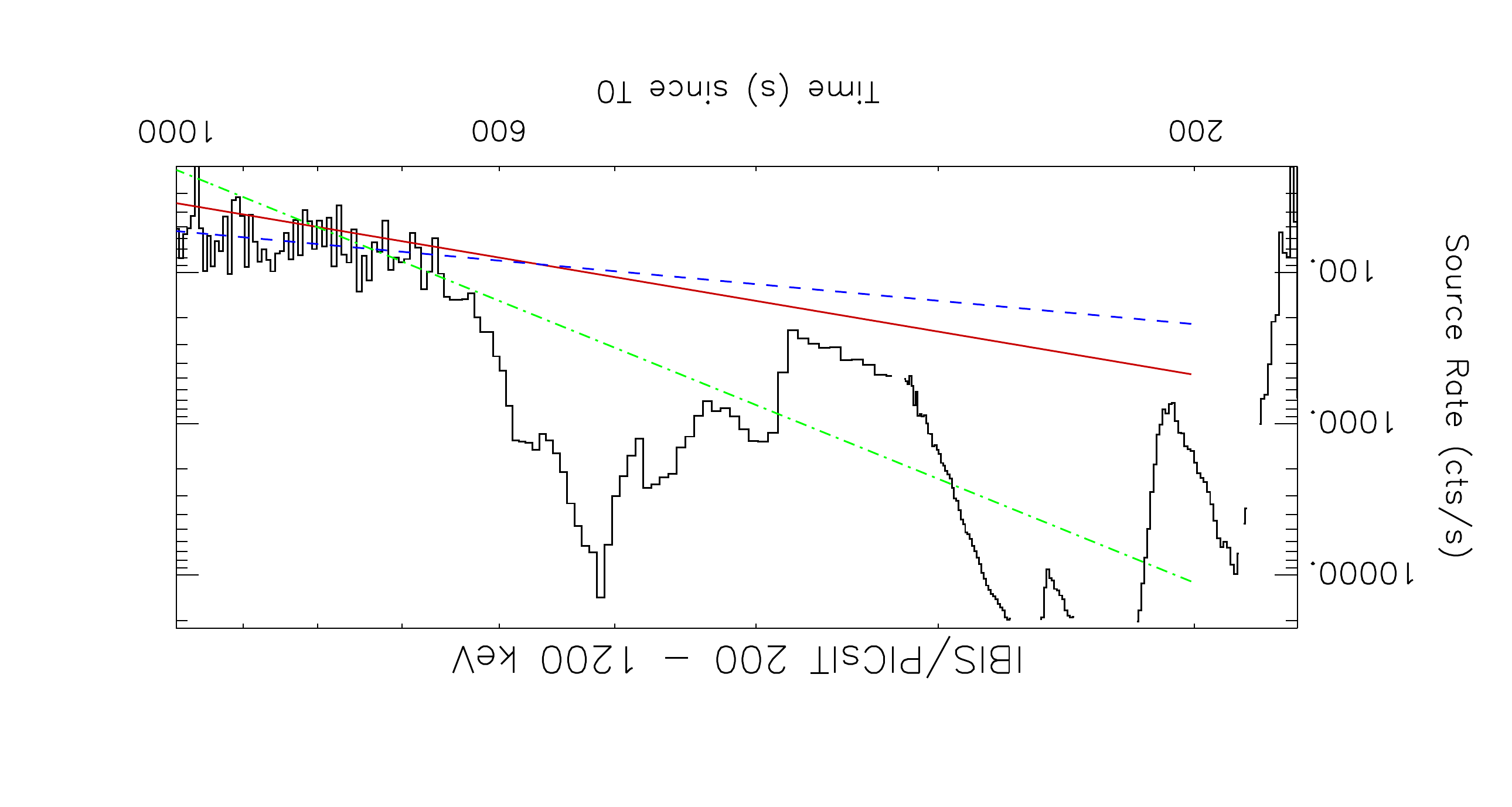} 
\caption{GRB 221009A \(200-1200\) keV \textrm{PICsIT} light curve. The power-law best-fit slope to the data from \(T_0 +600 -900\) s is a slope of 3.91 and is shown in green.  The best-fit to the data from  \( T_0 +630 - 900\) s gives an index of \( \ 1.6\), which is shown as the red line.  The blue dashed line is the best-fit a slope from \cite{an2023} for \(T_0 +600 - 900 \) in the \(600 - 3000\) keV energy range and is a slope of 0.88.}
    \label{fig:afterglow}
\end{figure*}

Detailed analysis of afterglow observations with \textit{INTEGRAL} data after \(T_0 + 1000\) s are covered in \citep{savchenko2023}.  However, \cite{an2023} report that afterglow emission begins after \(T_0 + 225\) s while the prompt emission is still present.  To search for the presence of afterglow emission prior to \(T_0 + 600\) s, we fit the \textrm{PICsIT} light curve to a power law (\(N(t) \propto t^{-\alpha}\)) from \(T_0 +600 - 900\) s, following \cite{an2023} and found a best-fit indxes of \( \alpha = 3.91 \pm 0.01\), which is shown as the green dot-dash line in Figure~\ref{fig:afterglow}.  This is significantly different than the \( 0.88\) from \cite{an2023} (dashed blue line in Figure~\ref{fig:afterglow}).

An inspection of the \textrm{PICsIT} light curve (Figure~\ref{fig:afterglow}) finds the \cite{an2023} slope begins to over-predict the observed flux at approximately \(T_0 + 900\) s.  In contrast, the \(3.91\) slope under-predicts the observed flux above \( \sim T_0 +800\) s, requiring a break to describe the decay at later times.  \cite{an2023} report a break at \(T_0 + 1125\) s, well before what is required for either slope based on the \textrm{PICsIT} data.  Additionally, the 3.91 slope significantly over-predicts the observed flux between \( \sim T_0 +300 -380\) s.  Thus a start time of \(T_0 + 600 \) s for when the afterglow emission begins to dominate is not consistent with the \textrm{PICsIT} data. 

The \textrm{PICsIT} light curve shows significantly different slopes at \(T_0 +600 \) s and \(T_0 + 700\)s indicating that the afterglow emission begins to dominate between these times.  While a precise start time of when the afterglow emission begins to dominate is difficult to determine using the \textrm{PICsIT} light curve, we tested several start times requiring that the extrapolated flux not exceed the observed flux prior to \(T_0 +225\) s.  We found \(T_0 +630 \) s as an approximate start time when the light curve decay has a slope of \( \alpha = 1.6 \pm 0.2\) (solid red line).  This slope is similar to the \(T_0+1350 -1860\) s fit of \( 1.89 \pm 0.07 \) from \cite{an2023} and consistent with the soft X-ray slopes of \( \sim 1.5\) \citep{oconnor2023,williams2023} and the soft gamma-ray slope (\(\sim 1.78\)) from \cite{savchenko2023}.

Spectral analysis in the \(600 -3000 \) keV energy range by \textit{Insight-HMXT} during \(T_0 + 630 -930\) s found a power-law spectrum with an index of \( 1.62\) using the \textrm{HE/CsI} and \textit{GECAM} \(> 200\) keV had a spectral index of \( \Gamma = 1.56 \pm 0.16 \) \citep{an2023}.  We found the \textrm{PICsIT} data during the same period are unable to constrain a photon index. Interestingly, the photon indexes are also consistent with the values from \textrm{PICsIT} during \(T_0 +400 -450 \) s and \(T_0 + 550 - 600\) s, surrounding Pulse 4 when the spectrum softens.  These spectra are also significantly harder than the spectra seen during Pulses 1, 2, and 3, suggesting the emission at these later, fainter periods are different from the brighter periods. 


\section{Conclusions}

In conclusion, the study of GRB 221009A spectral evolution in the \(200-2600\) keV energy range using the \textrm{PICsIT} spectral-timing data shows that the prompt emission has a "flux-tracking" behavior with the \textrm{PICsIT} power-law indexes evolving in a correlated way with the GRB flux.  Similar behavior was reported by \cite{yang2023}.  They interpret the spectral evolution as synchrotron emission with an expending emission region and a decaying global magnetic field without the need for photospheric emission.

An investigation of the photon index-flux correlation for each pulse and the rise and decay phase (where the data are present) showed that the relationship varies across pulses and sometimes between the rise and decay phases of the same pulse.  Additionally, the strength of the correlation appears to weaken as the prompt emission progresses with decays phases of Pulses 3 and 4 showing little to no correlation at low fluxes (\(< 0.25 \) ph/cm\(^2\)/s/keV).  

Lastly, we searched for the presence of afterglow emission in the \textrm{PICsIT} data prior to \(T_0 + 600\) s, finding the afterglow emission dominates after \( \sim T_0 +630\) s and decays with a slope of \( 1.6 \pm 0.2\) until at least \(T_0 +900 \) s.  Also, this decay index is consistent with those seen at soft X-rays prior to \(T_0 + 79000\) s.

\begin{acknowledgements}
      We thank the anonymous referee for their comments and suggestions.  The authors thank the Italian Space Agency for the financial support under the “INTEGRAL ASI-INAF” agreement n◦ 2019-35-HH.0.  The research leading to these results has received funding from the European Union’s Horizon 2020 Programme under the AHEAD2020 project (grant agreement n. 871158)
\end{acknowledgements}

%

\begin{thebibliography}{}



\bibitem[An et al.(2023)]{an2023} An, Z.-H., Antier, S., Bi, X.-Z., et al.\ 2023, arXiv:2303.01203. doi:10.48550/arXiv.2303.01203

\bibitem[Battiston et al.(2023)]{battiston2023} Battiston, R., Neub\"{u}ser,. C., Follega, F.~M., et al. 2023, ApJL, 964, L29. doi:10.3847/2041-8213/acc247


\bibitem[Bianchin et al.(2009)]{bianchin2009} Bianchin, V., Foschini, L., Di Cocco, G., et al.\ 2009, Advances in Space Research, 43, 1055. doi:10.1016/j.asr.2009.01.002

\bibitem[Bianchin et al.(2011)]{bianchin2011} Bianchin, V., Mereghetti, S., Guidorzi, C., et al.\ 2011, \aap, 536, A46. doi:10.1051/0004-6361/201117290

\bibitem[Bissaldi et al.(2022)]{bissaldi2022} Bissaldi, E., Omodei, N., Kerr, M., et al.\ 2022, GRB Coordinates Network, Circular Service, No. 32637, 32637

\bibitem[Burns et al.(2023)]{burns2023} Burns, E., Svinkin, D., Fenimore, E., et al.\ 2023, arXiv:2302.14037


\bibitem[Dichiara et al.(2022)]{dichiara2022} Dichiara, S., Gropp, J.~D., Kennea, J.~A., et al.\ 2022, The Astronomer's Telegram, 15650

\bibitem[ESA(2022)]{esa2022} \url{https://www.esa.int/ESA_Multimedia/Images/2022/10/ESA_spacecraft_catch_the_brightest_ever_gamma-ray_burst}

\bibitem[Frederiks et al.(2022)]{frederiks2022} Frederiks, D., Lysenko, A., Ridnaia, A., et al.\ 2022, GRB Coordinates Network, Circular Service, No. 32668, 32668

\bibitem[Frederiks et al.(2023)]{frederiks2023} Frederiks, D., Svinkin, D., Lysenko, A.~L., et al.\ 2023, arXiv:2302.13383. doi:10.48550/arXiv.2302.13383


\bibitem[Gotz et al.(2022)]{gotz2022} Gotz, D., Mereghetti, S., Savchenko, V., et al.\ 2022, GRB Coordinates Network, Circular Service, No. 32660, 32660


\bibitem[Jensen et al.(2003)]{jensen2003} Jensen, P. L., Clausen, K., Cassi, C. et al. 2003, A\&A, 411, L7

\bibitem[Labanti et al.(2003)]{labanti2003} Labanti, C., Di Cocco, G., Ferro, G., et al.\ 2003, \aap, 411, L149. doi:10.1051/0004-6361:20031356

\bibitem[Lapshov et al.(2022)]{lapshov2022} Lapshov, I., Molkov, S., Mereminsky, I., et al.\ 2022, GRB Coordinates Network, Circular Service, No. 32663, 32663

\bibitem[Li et al.(2021)]{li2021} Li, L., Ryde, F., Pe'er, A., et al.\ 2021, \apjs, 254, 35. doi:10.3847/1538-4365/abee2a


\bibitem[Liu et al.(2022)]{liu2022} Liu, J.~C., Zhang, Y.~Q., Xiong, S.~L., et al.\ 2022, GRB Coordinates Network, Circular Service, No. 32751, 32751



\bibitem[Mitchell et al.(2022)]{mitchell2022} Mitchell, L.~J., Phlips, B.~F., \& Johnson, W.~N.\ 2022, GRB Coordinates Network, Circular Service, No. 32746, 32746


\bibitem[Negoro et al.(2022)]{negaro2022} Negoro, H., Nakajima, M., Kobayashi, K., et al.\ 2022, The Astronomer's Telegram, 15651

\bibitem[O'Connor et al.(2023)]{oconnor2023} O'Connor, B., Troja, E., Ryan, G., et al.\ 2023, arXiv:2302.07906. doi:10.48550/arXiv.2302.07906


\bibitem[Piano et al.(2022)]{piano2022} Piano, G., Verrecchia, F., Bulgarelli, A., et al.\ 2022, GRB Coordinates Network, Circular Service, No. 32657, 32657

\bibitem[Ripa et al.(2022)]{ripa2022} Ripa, J., Pal, A., Werner, N., et al.\ 2022, GRB Coordinates Network, Circular Service, No. 32685, 32685

\bibitem[Ripa et al.(2023)]{ripa2023} Ripa, J., Takahashi, H., Fukazawa, Y., et al.\ 2023, arXiv:2302.10047. doi:10.48550/arXiv.2302.10047

\bibitem[Rodi et al.(2021)]{rodi2021} Rodi, J., Ubertini, P., Bazzano, A., et al.\ 2021, 43rd COSPAR Scientific Assembly. Held 28 January - 4 February

\bibitem[Savchenko et al.(2017)]{savchenko2017} Savchenko, V., Bazzano, A., Bozzo, E., et al.\ 2017, \aap, 603, A46. doi:10.1051/0004-6361/201730572

\bibitem[Savchenko et al.(2023)]{savchenko2023} Savchenko, V., Bazzano, A., Ubertini, P., et al. 2023, Submitted

\bibitem[Tan et al.(2022)]{tan2022} Tan, W.~J., Li, C.~K., Ge, M.~Y., et al.\ 2022, The Astronomer's Telegram, 15660

\bibitem[Ubertini et al.(2003)]{ubertini2003} Ubertini, P., Lebrun, F., Cocco, G. D., et al. 2003, 139

\bibitem[Ursi et al.(2022)]{ursi2022} Ursi, A., Panebianco, G., Pittori, C., et al.\ 2022, GRB Coordinates Network, Circular Service, No. 32650, 32650

\bibitem[Vedrenne et al.(2003)]{vedrenne2003} Vedrenne, G., Roques, J.-P., Schönfelder, V., et al. 2003, Astronomy \& Astrophysics, 411, L63 et al. 2003a, Astron- omy and Astrophysics, 411, L1


\bibitem[Veres et al.(2022)]{veres2022} Veres, P., Burns, E., Bissaldi, E., et al.\ 2022, GRB Coordinates Network, Circular Service, No. 32636, 32636

\bibitem[von Kienlin et al.(2003)]{vonkienlin2003} von Kienlin, A., Beckmann, V., Rau, A., et al.\ 2003, \aap, 411, L299. doi:10.1051/0004-6361:20031231

\bibitem[Williams et al.(2023)]{williams2023} Williams, M.~A., Kennea, J.~A., Dichiara, S., et al.\ 2023, arXiv:2302.03642. doi:10.48550/arXiv.2302.03642


\bibitem[Winkler et al.(2003)]{winkler2003} Winkler, C., Courvoisier, T. J.-L., Di Cocco, G., et al. 2003, Astron- omy and Astrophysics, 411, L1

\bibitem[Xiao et al.(2022)]{xiao2022} Xiao, H., Krucker, S., \& Daniel, R.\ 2022, GRB Coordinates Network, Circular Service, No. 32661, \#1 (2022/October-0), 32661

\bibitem[Yang et al.(2023)]{yang2023} Yang, J., Zhao, X.-H., Yan, Z., et al.\ 2023, arXiv:2303.00898. doi:10.48550/arXiv.2303.00898

 
\end{thebibliography}
%

\end{document}